\documentclass[referee]{raa}           
\usepackage{dcolumn}
\usepackage{bm}
\usepackage{graphicx,times}
\usepackage{amssymb,amsmath}

\usepackage{arydshln}
\usepackage[T1]{fontenc} 

\def\be{\begin{equation}}
\def\ee{\end{equation}}
\def\ba{\begin{eqnarray}}
\def\ea{\end{eqnarray}}

\def\ie{{\frenchspacing\it i.e.}}
\def\eg{{\frenchspacing\it e.g.}}

\def\ap{a_{\rm p}}
\def\dd{{\rm d}}
\def\lcdm{$\Lambda$CDM~}

\begin{document}

   \title{A new consistency test for \lcdm cosmology using galaxy surveys}

 \volnopage{ {\bf 2022} Vol.\ {\bf X} No. {\bf XX}, 000--000}
   \setcounter{page}{1}

   \author{Jinglan Zheng
   \inst{1,2,3}, Gong-Bo Zhao\inst{1,2}, Yuting Wang\inst{1},Xiaoyong Mu
      \inst{1,2},  Ruiyang Zhao\inst{1,2}, Weibing Zhang\inst{1,2}, Shuo Yuan\inst{1}, David Bacon\inst{3}, Kazuya Koyama\inst{3}
   }

   \institute{ National Astronomy Observatories, Chinese Academy of Sciences, Beijing, 100101, P.R.China; {\it jinglan.zheng@myport.ac.uk; gbzhao@nao.cas.cn}\\
        \and
             School of Astronomy and Space Science, University of Chinese Academy of Sciences, Beijing 100049, P.R.China\\ 
        \and 
            Institute of Cosmology and Gravitation, University of Portsmouth, Dennis Sciama Building, Portsmouth PO1 3FX, United Kingdom\\
\vs \no
}

\abstract{We propose a new consistency test for the $\Lambda$CDM cosmology using baryonic acoustic oscillations (BAO) and redshift space distortion (RSD) measurements from galaxy redshift surveys. Specifically, we determine the peak position of $f\sigma_8(z)$ in redshift $z$ offered by a RSD measurement, and compare it to the one predicted by the BAO observables assuming a flat $\Lambda$CDM cosmology. We demonstrate this new test using the simulated data for the DESI galaxy survey, and argue that this test complements those using the background observables alone, and is less subject to systematics in the RSD analysis, compared to traditional methods using values of $f\sigma_8(z)$ directly.
}

   \authorrunning{J. Zheng et al. }            
   \titlerunning{A new consistency test for \lcdm}  
   \maketitle

\section{Introduction}
\label{sec:intro}

The $\Lambda$CDM model, in which the cold dark matter (CDM) and a cosmological constant, $\Lambda$, contribute to roughly $1/3$ and $2/3$ energy budget of the current Universe respectively, has become the standard cosmological paradigm (\cite{R98,P99}). Although this `vanilla' model is favoured by most observations available so far in terms of model selection, it is being challenged, especially by the `Hubble crisis' (see \cite{H0T} for a recent review).   

Performing consistency tests for the $\Lambda$CDM model is one of the most efficient ways to discover new physics, if any, beyond the standard cosmological scenario, and efforts have been made along these lines. For example, the quantity $Om(z)$ (\cite{Om}), derived from $H(z)$ using cosmic chronometers measuring the age of passive galaxies at various redshifts (\cite{OHD}), complemented with the local $H_0$ measurement (\cite{R22}) can be used for a consistency test. $Om(z)$ is a constant and coincides with $\Omega_{\rm M,0}$ only if the underlying cosmology is $\Lambda$CDM, while it evolves with redshift otherwise. This quick test relies on measuring  both $H(z)$ and $H_0$ precisely, which is challenging. Further, this test only accounts for the expansion history of the Universe. 

In this paper, we propose a new consistency test for the $\Lambda$CDM model only using observables delivered by spectroscopic galaxy surveys; namely, we use baryonic acoustic oscillations (BAO) (\cite{BAO}) and redshift space distortion (RSD) (\cite{RSD}) measurements at multiple redshifts to hunt for deviations from the $\Lambda$CDM model at both the background and perturbation levels, and demonstrate our method using simulated Dark Energy Spectroscopic Instrument (DESI) measurements (\cite{DESI}). 

The new method is presented in Section \ref{sec:method}, including the relevant formalism and procedure; we then show the main result in Section \ref{sec:result}, before concluding in Section \ref{sec:conclusion}.

\section{Methodology}
\label{sec:method}

\subsection{The idea and procedure}

We start from the well-known relations for the evolution of the matter density parameter and growth of structure (\cite{Dodelson}), \ba   
\label{eq:OmegaM} \Omega_{\rm M}(a) &=&  \frac{\Omega_{\rm M,0}}{a^3}\left[\frac{H_0}{H(a)}\right]^2  \\
\label{eq:f} f(a) &\equiv& \frac{\dd {\rm log} \ \delta(a)}{\dd {\rm log} \  a} = \Omega_{\rm M}^\gamma(a) \\
\label{eq:s8} \sigma_8(a) &\propto&  \delta(a) \ea where $H$ denotes the expansion rate of the Universe, $\Omega_{\rm M}$ is the fractional matter density, $\delta$ is the overdensity of matter and $\sigma_8$ is the root-mean-square (rms) matter fluctuation on a scale of $8~h^{-1}{\rm Mpc}$. Symbols with a subscript $0$ mean quantities at redshift $0$, and Eq (\ref{eq:f}) is a reasonable approximation relating the expansion history with structure growth, through the growth index $\gamma$ (\cite{gamma}).

 Combining Eqs (\ref{eq:OmegaM}) - (\ref{eq:s8}), we obtain
 \ba \label{eq:fs8p} \frac{\left(f\sigma_8\right)'}{f\sigma_8} &=& \frac{1}{a}\left\{{\Omega_{\rm M}^{\gamma}(a)} - \gamma[1-2q(a)]\right\}   \ea in which the prime denotes the derivative with respect to the scale factor, and $q$ is the deceleration parameter defined as $q\equiv - \ddot{a}a/\dot{a}^2$,  where the dot is the derivative with respect to time.
 
 According to Eq (\ref{eq:fs8p}), $f\sigma_8(a)$ has a peak in $a$ (and in $z$) at a specific redshift, namely, $a=\ap$, if \ba {\Omega_{\rm M}^{\gamma}}(\ap)& = &\gamma\left[1-2q(\ap)\right] \nonumber \\
\label {eq:peakgen} & = & 3  \gamma   \left[1-\Omega_{\rm M}(\ap) \right] \ea where the last equation holds in the $\Lambda$CDM scenario. To investigate whether the peak of $f\sigma_8$ exists for a range of cosmologies, we allow the growth index $\gamma$, the equation of state of dark energy $w$ (assumed to be a constant), and the fractional matter density at current epoch $\Omega_{\rm M,0}$, to vary within a wide range. Fig. 1 shows two groups of curves of ${\Omega_{\rm M}^{\gamma}}$ and $\gamma(1-2q)$ as functions of the scale factor $a$, calculated using Eqs. (\ref{eq:OmegaM}) and (\ref{eq:f}) with $H_0= 67.4~{\rm km~s^{-1}~Mpc^{-1}},~\Omega_{\rm M,0}=0.31,\gamma=6/11$, which are the values favored by the Planck $2018$ observations in the \lcdm model (\cite{PLC18}).

The two groups of curves intersect in all cases, which means that the peak exists for all these cosmologies. This is confirmed by Fig \ref{fig:fs8}, where $f\sigma_8$ for various cosmologies is shown.

\begin{figure}[tbp]
\centering 
\hfill
\includegraphics[width=\textwidth]{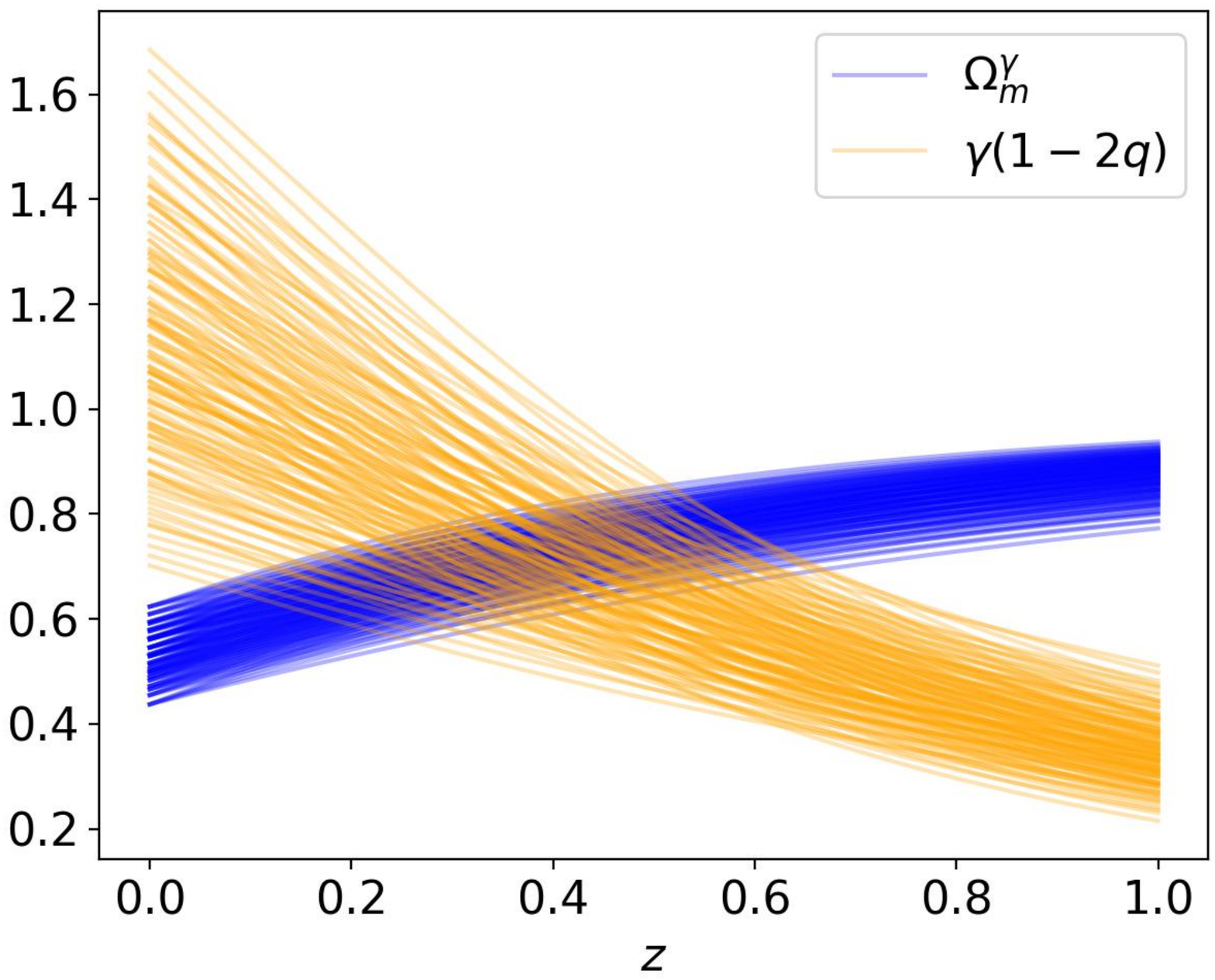}
\caption{\label{fig:peakequ} The orange and blue curves show quantities $\gamma (1-2q)$ and $\Omega_{\rm M} ^\gamma$ as a function of $a$, respectively. The collections of curves represent models with $5$ values of $w,\Omega_{\rm M,0}$ and $\gamma$ each, evenly distributed in ranges of $w \in [-1.2, -0.8]$ , $\Omega_{\rm M,0}  \in [0.28, 0.35]$ and $\gamma \in [0.45, 0.65]$, so that there are $125$ curves in total for both $\gamma (1-2q)$ and $\Omega_{\rm M} ^\gamma$.}
\label{intersect}
\end{figure}

\begin{figure}[htp]
\centering 
\hfill
\includegraphics[width=\textwidth]{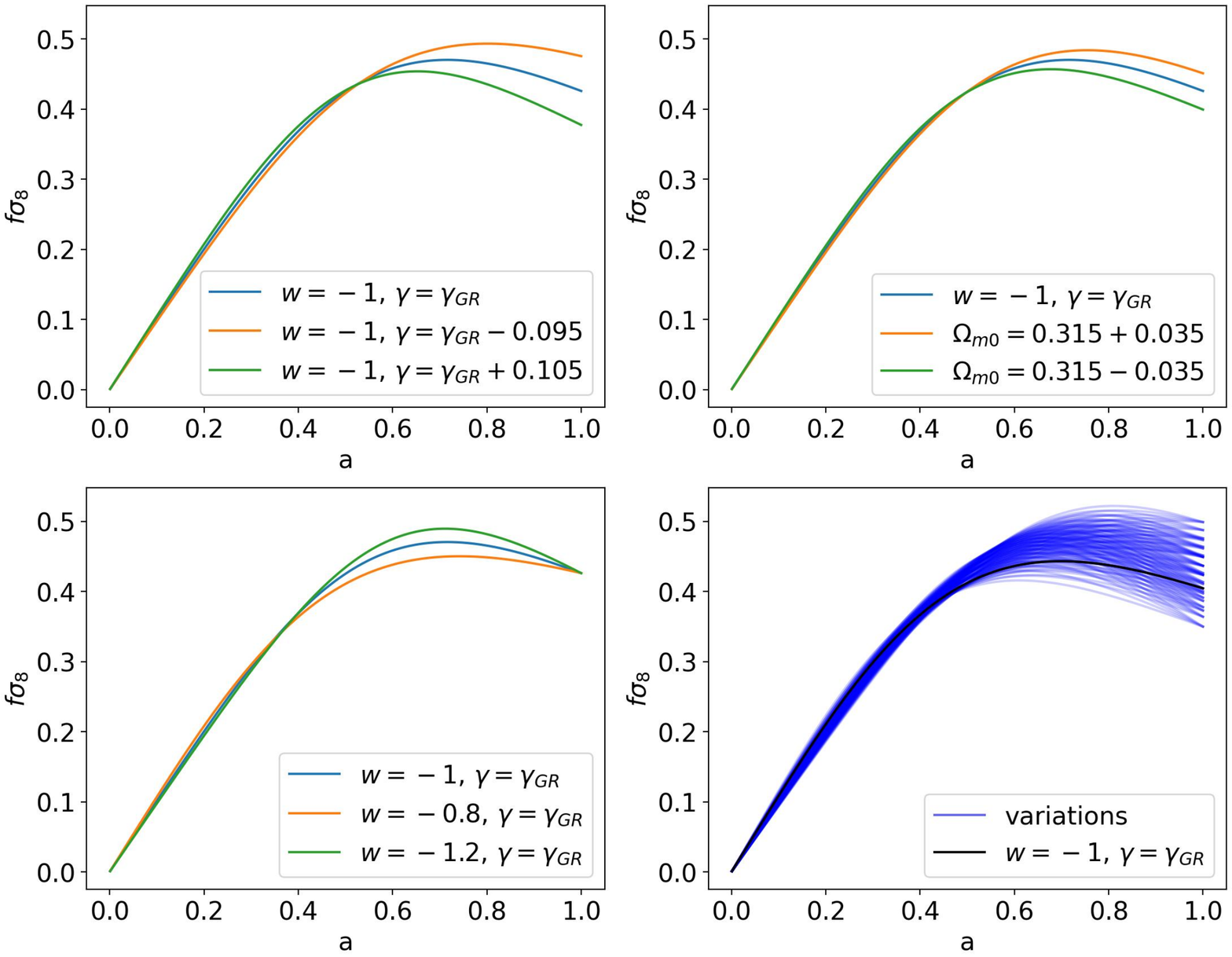}
\caption{\label{fig:fs8} The quantity $f\sigma_8$ as a function of $a$ for various cosmologies. In all panels, the fiducial $\Lambda$CDM model is shown in the middle for reference. `Variations' in the bottom right panel means $f\sigma_8$ for models ($125$ in total) with all parameters varied.}
\end{figure}

Given $\Omega_{\rm M,0}, w$ and $\gamma$, the position of the peak in $f\sigma_8(a)$ can be found by solving Eq (\ref{eq:peakgen}) explicitly. In $\Lambda$CDM, where $w=-1$ and $\gamma=6/11$ (\cite{gamma}), the peak position of $f\sigma_8(a)$ in $a$ is determined once $\Omega_{\rm M,0}$ is known. On the other hand, $\Omega_{\rm M,0}$ in a flat $\Lambda$CDM Universe can be found from $D_{\rm A}(z) H(z)$, the product of the radial and transverse distances at redshift $z$, which is provided by a BAO measurement. Note that in a flat $\Lambda$CDM Universe, $D_{\rm A}(z) H(z)$ only depends on $\Omega_{\rm M,0}$.

This motivates a new consistency test for $\Lambda$CDM model using BAO and RSD measurements, which are provided by galaxy surveys. The procedure for this new test is as follows:
\begin{enumerate}
\item Given a pair of measured $D_{\rm A}/r_{\rm d}$ and $H r_{\rm d}$ from BAO in galaxy surveys at a specific effective redshift, compute $D_{\rm A} H$, from which $\Omega_{\rm M,0}$ is computed assuming a $\Lambda$CDM model, denoted as $\Omega_{\rm M,0}^{\rm BAO}$;

\item Take measurements of $D_{\rm A} H$ available from BAO at other redshifts; repeat step $1$ to extract $\Omega_{\rm M,0}^{\rm BAO}$ for those redshifts;

\item Quantify the agreement of $\Omega_{\rm M,0}$ extracted from observables at various redshifts, by fitting a constant, denoted as $\bar{\Omega}_{\rm M,0}^{\rm BAO}$, to all the data points of $\Omega_{\rm M,0}^{\rm BAO}$;

\item Given $\bar{\Omega}_{\rm M,0}^{\rm BAO}$, compute the expected peak position of $f\sigma_8$ in $\Lambda$CDM, denoted as $z_{\rm p}^{\rm BAO}$, using Eq (\ref{eq:peakgen});

\item Using the measured $f\sigma_8$ data points, determine the actual peak position using a Taylor expansion approach (see next subsection) and denote this as $z_{\rm p}^{\rm RSD}$;

\item Compare $z_{\rm p}^{\rm BAO}$ with $z_{\rm p}^{\rm RSD}$.
\end{enumerate}

\noindent In summary, this new consistency test contains two key ingredients: testing the agreement among $\Omega_{\rm M,0}^{\rm BAO}$ derived at multiple redshifts, and the agreement between $z_{\rm p}^{\rm BAO}$ and $z_{\rm p}^{\rm RSD}$. To quantify the (dis)agreement, we compute two quantities as follows:
\begin{itemize}
	\item 
	
	(S/N)$_{\bf B}$: 
	This is to quantify the significance of ${\Omega}_{\rm M,0}^{\rm BAO}$ derived at various redshifts not being a constant. Practically, we fit a constant, which is $\bar{\Omega}_{\rm M,0}^{\rm BAO}$, to all ${\Omega}_{\rm M,0}^{\rm BAO}$ data points and record the $\chi^2$ for the best-fit value of $\bar{\Omega}_{\rm M,0}^{\rm BAO}$. The covariance between $D_A$ and $H$ is properly taken into account using a Fisher matrix analysis. Then (S/N)$_{\bf B}$ is defined as $\sqrt{\chi^2}$. 
	
	\item 
	(S/N)$_{\bf P}$:
	This is to quantify the agreement between $z_{\rm p}^{\rm BAO}$ and $z_{\rm p}^{\rm RSD}$, and is calculated as $|z_{\rm p}^{\rm BAO} - z_{\rm p}^{\rm RSD}|/\sigma$, where $\sigma$ is the uncertainty of $(z_{\rm p}^{\rm BAO} - z_{\rm p}^{\rm RSD})$; we discuss how this is evaluated in Section 2.2. We assume that measurements at the different redshifts have negligible covariance.
\end{itemize}

\begin{figure}[htp]
\centering 
\hfill
\includegraphics[width=0.9\textwidth]{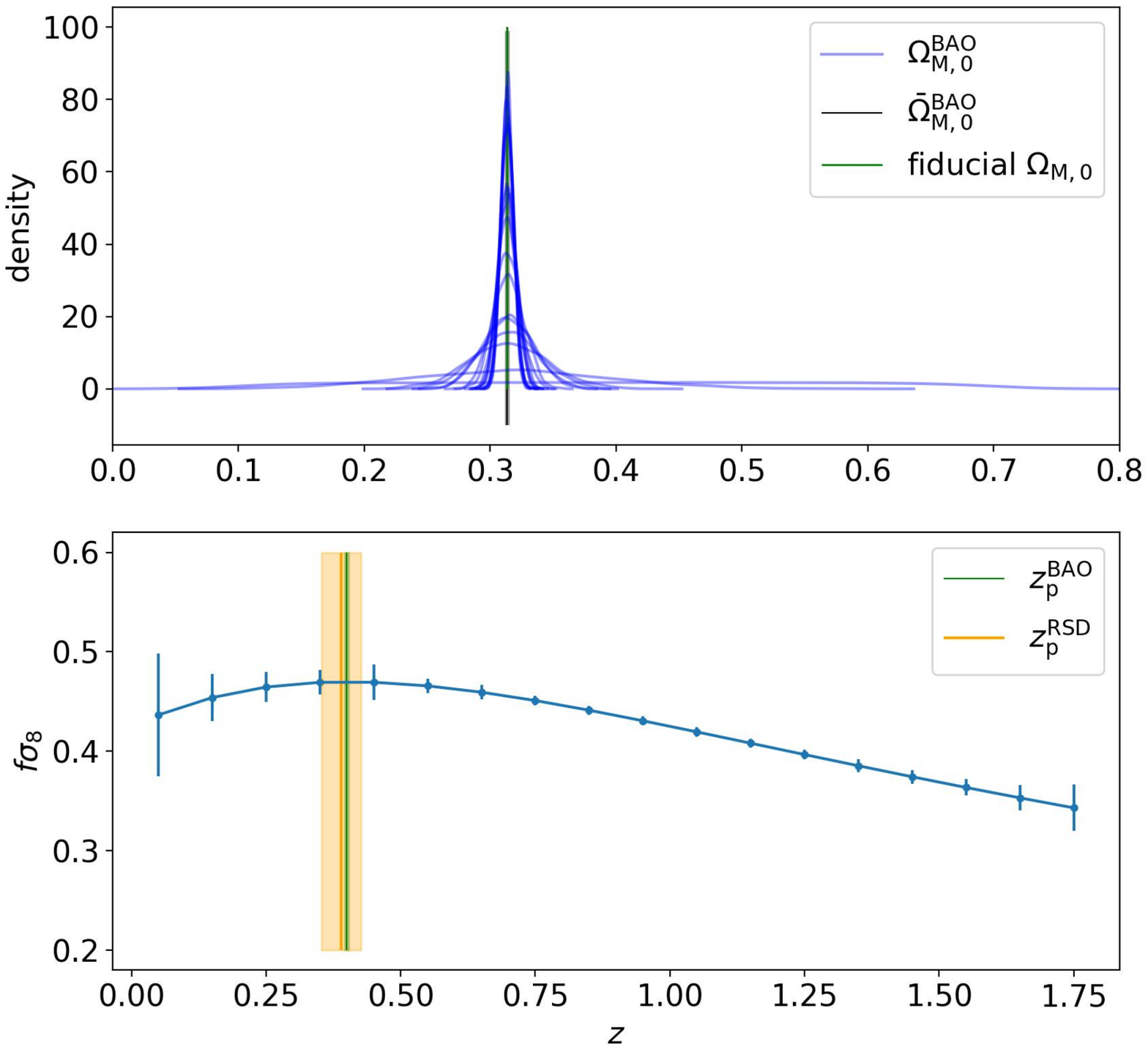}
\caption{\label{fig:modelI} The derived $\bar{\Omega}_{\rm M,0}^{\rm BAO}, z_{\rm p}^{\rm BAO}$ and $z_{\rm p}^{\rm RSD}$ for the $\Lambda$CDM model. Upper panel: The 1D distribution of ${\Omega}_{\rm M,0}^{\rm BAO}$ at $18$ redshifts (blue) and $\bar{\Omega}_{\rm M,0}^{\rm BAO}$ (black). The green vertical line denotes the fiducial ${\Omega}_{\rm M,0}$, which is consistent with the Planck 2018 cosmology used for the forecast. Lower panel: The forecast $f\sigma_8$ (blue data points with error bars), $z_{\rm p}^{\rm BAO}$ (thin green band) and $z_{\rm p}^{\rm RSD}$ (thick orange).}
\end{figure}

\begin{figure}[htp]
\centering 
\hfill
\includegraphics[width=0.9\textwidth]{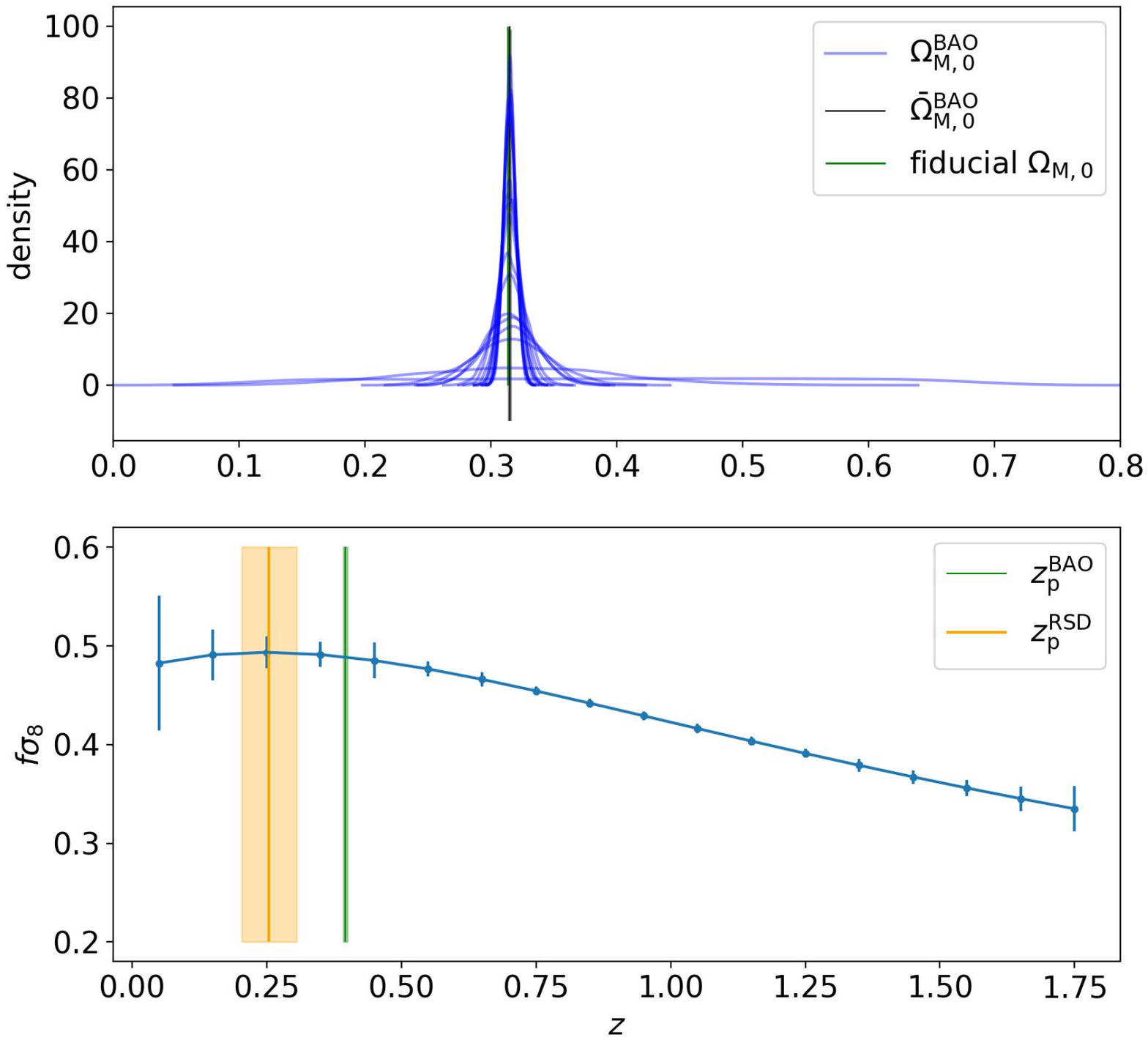}
\caption{\label{fig:modelII} Same as Fig. \ref{fig:modelI} but for model II ($w=-1, \  \gamma=0.45$.)}
\end{figure}

\begin{figure}[htp]
\centering 
\hfill
\includegraphics[width=0.9\textwidth]{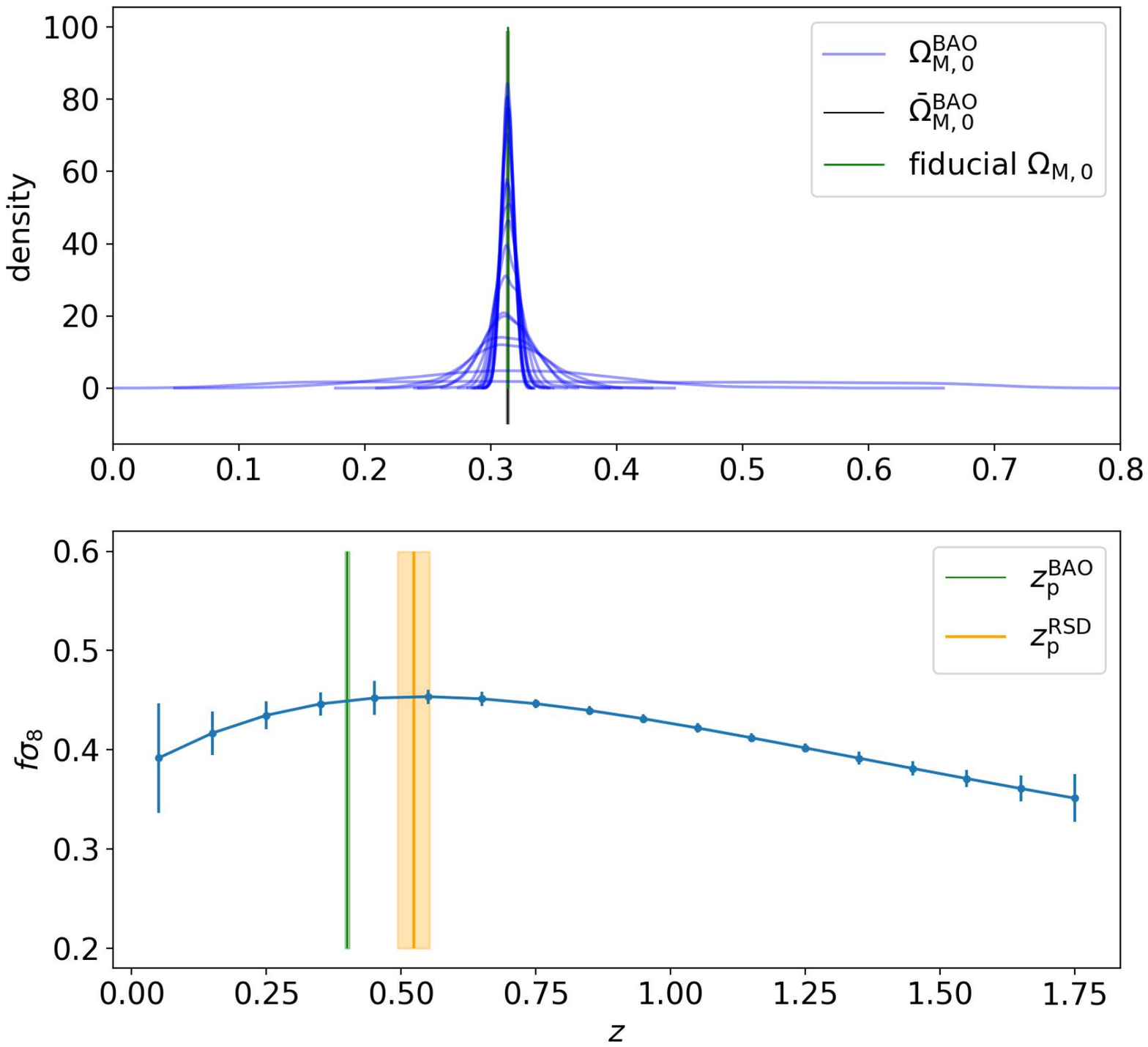}
\caption{\label{fig:modelIII} Same as Fig. \ref{fig:modelI} but for model III ($w=-1, \  \gamma=0.65$).}
\end{figure}

\begin{figure}[htp]
\centering 
\hfill
\includegraphics[width=0.9\textwidth]{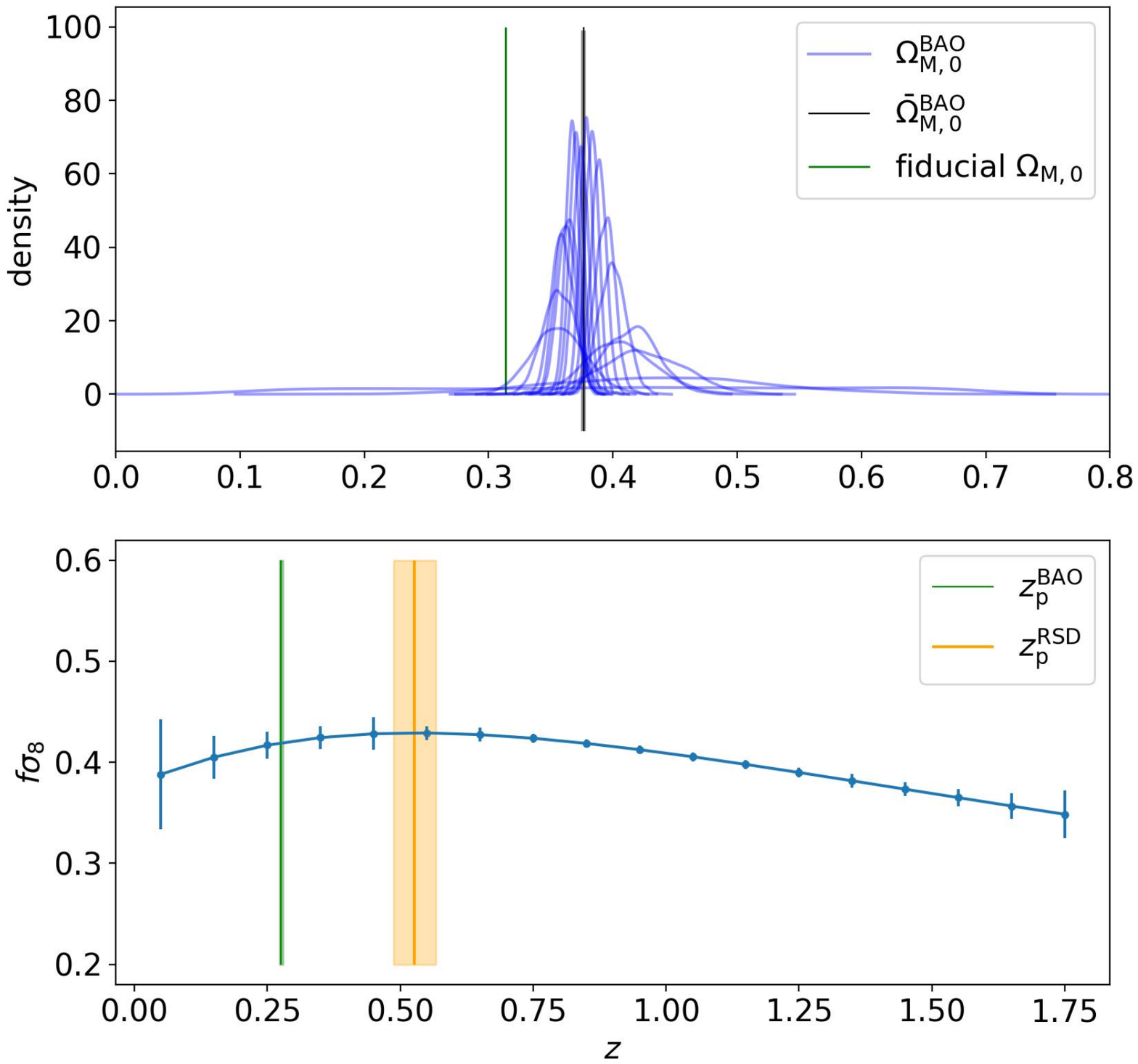}
\caption{\label{fig:modelIV} Same as Fig. \ref{fig:modelI} but for model IV ($w=-0.8, \  \gamma=0.65$). }
\end{figure}

\subsection{Determine $\bar{\Omega}_{\rm M,0}^{\rm BAO}, z_{\rm p}^{\rm BAO}$ and $z_{\rm p}^{\rm RSD}$}

The new consistency test requires measuring $\bar{\Omega}_{\rm M,0}^{\rm BAO}, z_{\rm p}^{\rm BAO}$ and $z_{\rm p}^{\rm RSD}$ from galaxy surveys. Here we describe how to measure these quantities from the simulated data, assuming a DESI sensitivity for a demonstration. 

Given a cosmological model, which in this work is considered to be $w\gamma$CDM (a CDM model with a constant $w$ for dark energy and a growth index $\gamma$ for matter in a flat Universe), we first perform a Fisher matrix forecast for $D_{\rm A}H$ and $f\sigma_8$ (with all relevant correlation coefficients) at $18$ effective redshifts (uniform in $z$ from $z=0.05$ to $1.75$) jointly covered by the LRGs and ELGs to be observed by a $14,000$ deg$^2$ DESI survey, with ${\rm d}N/{\rm d}z$  specified in Table 2.3 in the official DESI forecast paper (\cite{DESI}).

\begin{description}

\item[Determine $\bar{\Omega}_{\rm M,0}^{\rm BAO}$:] Given the uncertainties on $D_{\rm A}H$ at multiple redshifts ($18$ for the DESI example) obtained in the Fisher forecast, generate a large sample of $D_{\rm A}H$ at each redshift following a Gaussian distribution, with the mean and variance given by the fiducial model assumed and the Fisher forecast, respectively. This allows for a derivation of ${\Omega}_{\rm M,0}^{\rm BAO}$ from $D_{\rm A}H$ at each redshift assuming a flat $\Lambda$CDM cosmology. Then fit a constant, which is $\bar{\Omega}_{\rm M,0}^{\rm BAO}$, to the derived collection of ${\Omega}_{\rm M,0}^{\rm BAO}$ at all redshifts. The fitting is performed using \texttt{zeus} (\cite{zeus}), the Python package for slice sampling.

\item[Determine $z_{\rm p}^{\rm BAO}$:] Given the posterior of $\bar{\Omega}_{\rm M,0}^{\rm BAO}$ obtained from the previous step, draw a large sample of $\bar{\Omega}_{\rm M,0}^{\rm BAO}$, and solve for the peak location using the peak equation, Eq. (\ref{eq:peakgen}) for each $\bar{\Omega}_{\rm M,0}^{\rm BAO}$ in the sample using $\gamma=6/11$ (i.e. assuming GR); then compute the mean and uncertainty of $z_{\rm p}^{\rm BAO}$ accordingly.

\item[Determine $z_{\rm p}^{\rm RSD}$:] To determine $z_{\rm p}^{\rm RSD}$ from the simulated RSD measurements, we choose to use a cubic function to fit to a collection of simulated $f\sigma_8$ data points. The fitting function takes the form \ba f{\sigma_8}(a) = A + B\left(a - {\ap}\right)^2 + C\left(a - {\ap}\right)^3\ea where $A=f{\sigma_8}({\ap})$, $B=f{\sigma_8}^{\prime \prime}({\ap})/2$, $C=f{\sigma_8}^{\prime \prime \prime}({\ap})/6$ and $z_{\rm p}^{\rm RSD}=1/\ap-1$. This is a Taylor expansion of $f{\sigma_8}(a)$ around $\ap$, so the linear term vanishes by definition. We have performed tests on simulated data to confirm that this peak finder is sufficiently accurate given the DESI sensitivity (the bias is less than $4$\% in estimating the peak position in all cases).

\end{description}

\begin{table}[htp]
\centering
\begin{tabular}{c|c|c} 
\hline\hline
Models                & (S/N)$_{\bf B}$ & (S/N)$_{\bf P}$ \\ 
\hline
$w=-1, \ \gamma=\gamma_{\rm GR}$ & $0$ & $0.41$ \\
$w=-1, \ \gamma=\gamma_{\rm GR}-0.095$ &  $0$ & $2.43$\\
$w=-1, \ \gamma=\gamma_{\rm GR}+0.105$ &  $0$ & $4.14$\\
\hdashline 
$w=-1.2, \ \gamma=\gamma_{\rm GR}$          & $8.47$      & $3.86$   \\
$w=-1.2, \ \gamma=\gamma_{\rm GR}-0.095$ & $8.47$     & $4.71$   \\
$w=-1.2, \ \gamma=\gamma_{\rm GR}+0.105$ & $8.47$      & $0.23$    \\
\hdashline
$w=-0.8, \ \gamma=\gamma_{\rm GR}$       & $7.36$      & $1.47$  \\
$w=-0.8, \ \gamma=\gamma_{\rm GR}-0.095$ & $7.36$      & $1.96$    \\ 
$w=-0.8, \ \gamma=\gamma_{\rm GR}+0.105$ & $7.36$      & $6.30$   \\
\hline\hline
\end{tabular}
\caption{The significance of ${\Omega}_{\rm M,0}^{\rm BAO}$ derived from different redshifts not being a constant (middle column), and of $z_{\rm p}^{\rm BAO} \ne z_{\rm p}^{\rm RSD}$ (right column) for various models as listed.}
\label{tab}
\end{table}

\section{Results}
\label{sec:result}

In this section, we demonstrate the proposed new consistency test using the simulated BAO and RSD data assuming a DESI sensitivity. We perform the new consistency test on four different models: 
\begin{itemize}
	\item Model I: $w=-1, \ \gamma=\gamma_{\rm GR}=6/11$ ($\Lambda$CDM) 
	\item Model II: $w=-1, \ \gamma=\gamma_{\rm GR}-0.095=0.45$
	\item Model III: $w=-1, \ \gamma=\gamma_{\rm GR}+0.105=0.65$
	\item Model IV: $w=-0.8, \ \gamma=\gamma_{\rm GR}+0.105=0.65$
\end{itemize}

\noindent We carry out all of the procedures described in Section 2 for these models; the results are summarised in Figs. \ref{fig:modelI} - \ref{fig:modelIV} and Table \ref{tab}.

Fig. \ref{fig:modelI} shows the test result for the $\Lambda$CDM model. The upper panel shows the derived ${\Omega}_{\rm M,0}^{\rm BAO}$ from the simulated DESI $D_{\rm A} H$ measurement at $18$ effective redshifts, together with $\bar{\Omega}_{\rm M,0}^{\rm BAO}$, which is the compressed quantity from all the ${\Omega}_{\rm M,0}^{\rm BAO}$ data points. All ${\Omega}_{\rm M,0}^{\rm BAO}$ and $\bar{\Omega}_{\rm M,0}^{\rm BAO}$ agree with each other as expected, confirming that the model being tested is consistent with $\Lambda$CDM at the background level. On the other hand, the lower panel compares $z_{\rm p}^{\rm BAO}$ with  $z_{\rm p}^{\rm RSD}$, and an excellent agreement is reached, which further confirms that the model being tested is also consistent with $\Lambda$CDM at the perturbation level.

In Figs. \ref{fig:modelII} and \ref{fig:modelIII}, we show results for the cases of $w=-1, \ \gamma=0.45$ and $w=-1, \ \gamma=0.65$, respectively. In both cases, we see a discrepancy (around $2\sigma$ and $4\sigma$) between $z_{\rm p}^{\rm BAO}$ and $z_{\rm p}^{\rm RSD}$ in the lower panel. We also tried testing on models in which both expansion and growth history deviate from those in $\Lambda$CDM, \eg, $w=-0.8, \ \gamma=0.65$. As shown in Fig. \ref{fig:modelIV}, ${\Omega}_{\rm M,0}^{\rm BAO}$ at various redshifts do not agree (with a discrepancy around $\sim7.4\sigma$), and $z_{\rm p}^{\rm BAO}$ and $z_{\rm p}^{\rm RSD}$ differ at $\sim6.3\sigma$.

More test results are shown in Table \ref{tab}. Generally speaking, $z_{\rm p}^{\rm BAO}$ and $z_{\rm p}^{\rm RSD}$ are different if either $w\ne-1$, or $\gamma\ne\gamma_{\rm GR}$, making the new peak statistic  highly complementary to that for the background, \eg~the $\bar{\Omega}_{\rm M,0}^{\rm BAO}$ test or the $Om$ test (\cite{Om}). For models where $w\ne-1$ and $\gamma\ne\gamma_{\rm GR}$, $z_{\rm p}^{\rm BAO}$ and $z_{\rm p}^{\rm RSD}$ are also distinct if $w>-1, \ \gamma>\gamma_{\rm GR}$ or $w<-1, \ \gamma<\gamma_{\rm GR}$. For example, (S/N)$_{\bf P}$ can reach $4.71$ and $6.3\sigma$ level for $w=-1.2, \ \gamma=0.45$ and $w=-0.8, \ \gamma=0.65$, respectively. For models where both $w$ and $\gamma$ deviate from those in $\Lambda$CDM but in opposite directions, \eg, when $w>-1, \ \gamma<\gamma_{\rm GR}$ or $w<-1, \ \gamma>\gamma_{\rm GR}$, $z_{\rm p}^{\rm BAO}$ and $z_{\rm p}^{\rm RSD}$ can approach each other, because of the degeneracy between $w$ and $\gamma$ given the peak position. This could actually be used as a diagnosis for the model, \ie, a large (S/N)$_{\bf B}$ with a small (S/N)$_{\bf P}$ may suggest a deviation from $\Lambda$CDM model at both the background and perturbation level. The cosmological implication given  (S/N)$_{\bf B}$ and  (S/N)$_{\bf P}$ is summarised in Table \ref{tab:implication}.

\begin{table}[htp]
\centering
\begin{tabular}{c|c|c} 
\hline\hline
 (S/N)$_{\bf B}$  &  (S/N)$_{\bf P}$  & Implication\\ 
\hline
$\sim0$           &  $\sim0$          & $\Lambda$CDM \\
$\sim0$           &  Large          & Growth history deviates from that in $\Lambda$CDM \\
Large             &  $\sim0$ or large           & Expansion history deviates from that in $\Lambda$CDM; 
\\
&&growth may or may not deviate\\

\hline\hline
\end{tabular}
\caption{The cosmological implication given  (S/N)$_{\bf B}$ and  (S/N)$_{\bf P}$.}
\label{tab:implication}
\end{table}

\section{Discussion and Conclusions}
\label{sec:conclusion}
In this work we propose a new consistency test for the standard $\Lambda$CDM paradigm using the BAO and RSD measurements directly accessible from galaxy redshift surveys. 

This new test contains two essential ingredients: a test for the expansion history and a test for the structure growth. Consistency tests for the expansion history have been proposed in the literature, \eg, the $Om$ statistic, which also checks the constancy of $\Omega_{\rm M,0}$ derived from observables at different redshifts. However, the key difference between $Om$ and our test is that $Om$ relies on measurements of $H(z)$ and $H_0$, while ours only requires the BAO measurement. Direct $H(z)$ measurements are performed using the age of passive galaxies, and may be subject to large statistical and systematical uncertainties. The local $H_0$ measurement, on the other hand, is in serious tension with the indirect inference from the CMB, which may suggest new physics beyond $\Lambda$CDM, or unknown systematics. This makes our new test for the background more robust - the BAO measurements are known to be less contaminated by systematics (\cite{BOSSsys}), and are easier to access from existing galaxy surveys, including 2dFGRS (\cite{2dF}), SDSS-III BOSS (\cite{BOSS}), SDSS-IV eBOSS (\cite{eBOSS}), WiggleZ (\cite{WiggleZ}) and ongoing galaxy surveys such as DESI (\cite{DESI}), PFS (\cite{PFS}) and the Euclid mission (\cite{Euclid}). 

Our new test for the structure growth essentially scrutinises the consistency relation between the background and perturbation in the flat $\Lambda$CDM Universe. Since we only examine the peak position of $f\sigma_8$, we argue that this is less subject to possible systematics in the RSD measurement - the absolute value of $f\sigma_8$ is not directly used in this test. We note that the dominating error budget of $|z_{\rm p}^{\rm BAO} - z_{\rm p}^{\rm RSD}|$ is the uncertainty in $z_{\rm p}^{\rm RSD}$, which is primarily due to the low redshift resolution of $f\sigma_8$ measurements in traditional analyses. Recent developments, including the optimal redshift weighting method for RSD measurements (\cite{Zhao18,RR18}), may improve the $z_{\rm p}^{\rm RSD}$ measurement, which will be presented in a follow-up work using existing observations.

As we have access to observational data from stage-IV galaxy surveys in the near future, the new tests proposed in this work allow for an imminent and robust consistency check of the flat $\Lambda$CDM model.

\normalem
\begin{acknowledgements}
JZ is supported by the Chinese Scholarship Council and STFC. JZ, XM, RZ, WZ and GBZ are supported by NSFC Grants 11925303, 11720101004, 11890691. YW is supported by NSFC Grant 11890691, by the Youth Innovation Promotion Association CAS, and by the Nebula Talents Program of
NAOC. GBZ is also supported by the National Key Basic Research and Development Program of China, and a grant of CAS Interdisciplinary Innovation Team. We also acknowledge the science research grants from the China Manned Space Project with NO. CMS-CSST-2021-B01.

\end{acknowledgements}

\newpage
\appendix

\section{Derivation of Eq. (4)}

Let us start from

\ba f{\sigma_8} =  Af\delta \ea Then
\ba (f{\sigma_8})' = A(f\delta )' = Af\delta \left(\frac{f}{a} + \frac{{f'}}{f}\right) 
\ea where $' \equiv \frac{d}{{da}}$. 
Thus
\ba  
\frac{(f\sigma_8)^\prime}{f\sigma_8}= \frac{f}{a} + \frac{{f'}}{f}
\label{eq:fsp ratio}
\ea
Note that $f = \Omega_{\rm M}^\gamma$ and ${\Omega _{\rm M}}(a) = {\Omega _{\rm M,0}}{a^{ - 3}}{(H_0/H)^2}$, so we have

\ba \label{eq:f ratio 1} \frac{f^\prime}{f} &=& \gamma \frac{\Omega_{\rm M}^\prime}{\Omega_{\rm M}} \\
\frac{{{{\Omega'}_{\rm M}}}}{{{\Omega _{\rm M}}}} &=&  - \frac{3}{a} - 2\frac{{H'}}{H} =  - \frac{3}{a} + 2\frac{{1 + q}}{a} = \frac{{2q - 1}}{a} \label{eq:om ratio}
\ea where $q(a)  \equiv   - \frac{{\ddot aa}}{{\dot a}^2}$ is the deceleration parameter. This naturally leads to
\ba 
f = a\frac{{(f{\sigma _8})'}}{{f{\sigma _8}}} + \gamma (1 - 2q) 
\label{eq:fs8p2}
\ea which is exactly Eq. (\ref{eq:fs8p}).

\end{document}